\documentclass[preprint,draft,prl,10pt,twocolumn]{revtex4}

\input{tcilatex}

\begin{document}

\title{Experimental violation of a cluster state Bell inequality}
\author{Philip Walther$^{1}$, \ Markus Aspelmeyer$^{1}$, Kevin J. Resch$^{1}$
and Anton Zeilinger$^{1,2}$}
\affiliation{$^{1}$Institut f\"{u}r Experimentalphysik, Universit\"{a}t Wien,
Boltzmanngasse 5, A-1090 Wien, Austria\\
$^{2}$IQOQI, Institut f\"{u}r Quantenoptik und Quanteninformation, \"{O}%
sterreichische Akademie der Wissenschaften, Boltzmanngasse 3, A-1090 Wien,
Austria}

\begin{abstract}
Cluster states are a new type of multiqubit entangled states with
entanglement properties exceptionally well suited for quantum
computation. \ In the present work, we experimentally demonstrate
that correlations in a four-qubit linear cluster state cannot be
described by local realism.\ This exploration is based on a
recently derived Bell-type inequality [V. Scarani et al., Phys.
Rev A,  \textbf{71} 042325 (2005)] which is tailored, by using a
combination of three- and four-particle correlations, to be
maximally violated by cluster states but not violated at all by
GHZ states. \ We observe a cluster state Bell parameter of
$2.59\pm 0.08$, which is more than 7$\sigma $ larger than the
threshold of 2 imposed by local realism.
\end{abstract}

\maketitle

Multi-particle entanglement is a complex, and relatively unexplored
landscape. \ For two qubits, there exists only one class of entanglement\ %
\cite{Popescu}, for three qubits there are two classes of genuine
three-particle entangled states \cite{Wcirac,acin}, and for four qubits at
least nine different classes of entanglement have been identified \cite%
{verstraete}. Recently, a great deal of attention has been
devoted to a class of multiparticle entangled states called
cluster states. \ This attention is largely due to the
application of cluster states in Raussendorf and Briegel's
``one-way'' model for universal quantum computation
\cite{briegel}. \ In that model, one can drive a quantum
computation entirely through single-qubit measurements and
feedforward instead of unitary evolution. \ In addition to being
a practical alternative to the standard model for quantum
computing, it has also called into question the requirements for
quantum computing and the relationship between measurement and
dynamics \cite{Nielsen}.\ One-way quantum computation based on
cluster states demonstrating one-qubit gates, two-qubit gates,
and a
quantum search algorithm was recently realized experimentally \cite%
{clusterexp}.

Aside from their fascinating use for quantum computing, cluster states are a
novel kind of multiparticle entangled states with fundamentally new and
different properties. They share some properties with multi-particle
extensions of both Greenberger-Horne-Zeilinger (GHZ)\ states $\left| \mathrm{%
GHZ}\right\rangle =1/\sqrt{2}\left( \left| 000\right\rangle _{123}+\left|
111\right\rangle _{123}\right) $ \cite{GHZ,GHZbouw,GHZpan} and W states $%
\left| \mathrm{W}\right\rangle =1/\sqrt{3}\left( \left| 100\right\rangle
_{123}+\left| 010\right\rangle _{123}+\left| 001\right\rangle _{123}\right) $%
\cite{W,Wcirac,Wexp}. Each single-qubit constituent of a cluster state is
completely mixed, characteristic of GHZ states. Also, any two of the four
cluster-qubits can be projected into a Bell state by chosing an appropriate
basis, similar to a GHZ state, but cluster states also share their \emph{%
persistency of entanglement} \cite{Persistency} with the W states. Recent
theoretical investigations of the ``nonlocality'' of these cluster states
have constructed new types of Bell inequalities and even GHZ-type arguments
to refute local realism with the specific correlations of cluster states in
mind~\cite{scarani,briegelnonlocal}.

Bell's inequalities are specifically designed to put quantum physics to the
test against local realistic models. For two-qubit entangled states, the
CHSH-Bell inequality \cite{Bell,CHSH} is perhaps the best-known example.
The inequality is constructed from two-qubit spin or, in our case,
polarization correlation functions. \ Similarly, the Mermin inequality \cite%
{mermin}, testing  local realism in three-qubit entangled states, is made
entirely of three-qubit correlations. \ Its generalization is based entirely
on N-qubit correlations~\cite{zuk03}.  In general, the choice of these
correlations determines the optimality of a Bell inequality, i.e. whether
entanglement is detected by a maximal violation of the inequality. \ For
example, for the specific case of three qubits the inclusion of lower-order
correlations can lead to an optimal Bell inequality for a W state, which
could not detect GHZ entanglement~\cite{cabello}. \ This ambiguity and
selectivity of which type of entangled state produces a maximal Bell
violation makes the connection between entanglement and Bell's inequality
tenuous especially in multi-particle states \cite{gisinnonlocal} \
Nevertheless, it nicely highlights the fundamentally different ways in which
the GHZ and W states manifest violations of local realism. Since the number
of distinct classes of entanglement grows rapidly with the number of qubits,
one might expect to find other Bell inequalities optimal for different
states. Specifically, for a Bell inequality optimal for cluster states,
lower-order correlations will be of importance, since cluster states can be
generated by nearest-neighbor Ising interaction~\cite{briegel}.

A recent theoretical work has found a GHZ-type argument for cluster states %
\cite{scarani}. \ As in the original GHZ article \cite{GHZ}, the new work
showed that there exists a combination of observables whose expectation
values cannot be consistent with a set of local realistic properties.
However, in contrast to GHZ states, cluster states can even fulfill a GHZ
argument using combinations of three- and four-qubit correlations~\cite%
{footnote}{.} \ This leads to the development of a Bell-inequality that can
be maximally violated by cluster states but cannot be violated at all by
GHZ\ states. In this experimental work, we use four-qubit cluster states
encoded into the polarization state of photons to test that Bell inequality.

A linear cluster state arises when a line of qubits, each in the $\left|
+\right\rangle $ state, where $\left| \pm \right\rangle =\frac{1}{\sqrt{2}}%
\left( \left| 0\right\rangle \pm \left| 1\right\rangle \right) $, experience
nearest-neighbour CPhase operations, i.e., $\left| j\right\rangle \left|
k\right\rangle \rightarrow \left( -1\right) ^{jk}\left| j\right\rangle
\left| k\right\rangle ,$ $j,k\in \left\{ 0,1\right\} $\cite{briegel}. \
Linear cluster states of two and three qubits are equivalent under local
unitary transformations, or ``locally-equivalent'', to Bell states and GHZ
states, respectively \cite{briegel}. In contrast, the four-qubit cluster is
not locally equivalent to either the four-qubit GHZ\ or W states. \ In the
present work, we use photon polarization to encode qubits with horizontal
(vertical) polarization corresponding to $\left| 0\right\rangle $ $\left(
\left| 1\right\rangle \right) $. \ Our target cluster state is of the form,%
\begin{eqnarray}
\left| \phi _{4}\right\rangle  &=&\frac{1}{2}(\left| HHHH\right\rangle
_{1234}+\left| HHVV\right\rangle _{1234}+  \nonumber \\
&&\left| VVHH\right\rangle _{1234}-\left| VVVV\right\rangle _{1234}),
\end{eqnarray}%
where the subscripts 1,2,3 and 4 label different photons in separated
spatial modes. This state is locally-equivalent, under a Hadamard
transformation $H=\frac{1}{\sqrt{2}}\left( \sigma _{X}+\sigma _{Z}\right) $
on the first and last qubit, to the four-qubit linear cluster state
\begin{eqnarray}
\left| \phi _{4}^{\prime }\right\rangle  &=&\frac{1}{2}(\left|
0+0+\right\rangle _{1234}+\left| 0-1-\right\rangle _{1234}+  \nonumber \\
&&\left| 1-0+\right\rangle _{1234}+\left| 1+1-\right\rangle _{1234}),
\end{eqnarray}%
where $\left| \pm \right\rangle =\frac{1}{\sqrt{2}}\left( \left|
0\right\rangle \pm \left| 1\right\rangle \right) $ represents the
complementary linear polarization. The linear cluster state, $\left| \phi
_{4}^{\prime }\right\rangle $, has a set of 15 nontrivial stabilizer
operators, $S_{i}^{\prime }$, each made up of products of four Pauli
operators such that $S_{i}^{\prime }\left| \phi _{4}^{\prime }\right\rangle
=\pm \left| \phi _{4}^{\prime }\right\rangle $\cite{scarani}. \ Since each
of the Pauli operators, $\sigma _{X},$ $\sigma _{Y},$ $\sigma _{Z},$and $%
\sigma _{0}$, has eigenvalues of $\pm 1$ ($\sigma _{0}$ is the identity),
each such stabilizer operators represent a property of the state that is
fulfilled \emph{with certainty}, i.e. an element of physical reality~\cite%
{EPR}. \ Following the reasoning of GHZ one can then find sets of 4 of these
stabilizers, e.g. $\sigma _{Z}\sigma _{Y}\sigma _{Y}\sigma _{Z},$ $\sigma
_{Z}\sigma _{Y}\sigma _{X}\sigma _{Y},$ $\sigma _{0}\sigma _{Z}\sigma
_{X}\sigma _{Z}$, $\sigma _{0}\sigma _{Z}\sigma _{Y}\sigma _{Y}$ with
expectation values $+1$, $-1$, $+1$, and $+1$, which are inconsistent with
local realism. In addition, these stabilizers can be used to construct a
Bell inequality. \ Since $\left| \phi _{4}^{\prime }\right\rangle $ and $%
\left| \phi _{4}\right\rangle $ are equivalent only up to local
transformations, the stabilizer operators required for the GHZ argument need
to be interconverted. Obviously, the GHZ argument and Bell inequality,
remain intact. \ Making use of the relations $H\sigma _{X}=\sigma _{Z}H$, $%
H\sigma _{Z}=\sigma _{X}H$, $H\sigma _{Y}=-\sigma _{Y}H$, and $H\sigma
_{0}=\sigma _{0}H$, we can convert the four operators, $S_{i}^{\prime },$ to
a new set, $S_{i},$ to $\sigma _{X}\sigma _{Y}\sigma _{Y}\sigma _{X},$ $%
\sigma _{X}\sigma _{Y}\sigma _{X}\sigma _{Y},$ $\sigma _{0}\sigma _{Z}\sigma
_{X}\sigma _{X},$ and $\sigma _{0}\sigma _{Z}\sigma _{Y}\sigma _{Y}$, where
the expectation values for $\left| \phi _{4}\right\rangle $ are $+1$, $+1$, $%
+1$, $\ $and $-1$, respectively. \ These stabilizers can now be used to
construct the Bell inequality optimized for our cluster state $\left| \phi
_{4}\right\rangle .$ \ The Bell parameter, $S_{C}$, is given by%
\begin{eqnarray}
S_{C} &=&|\sigma _{X}\sigma _{Y}\sigma _{Y}\sigma _{X}+\sigma _{X}\sigma
_{Y}\sigma _{X}\sigma _{Y}|+  \nonumber \\
&&|\sigma _{0}\sigma _{Z}\sigma _{X}\sigma _{X}-\sigma _{0}\sigma _{Z}\sigma
_{Y}\sigma _{Y}|
\end{eqnarray}%
The assumptions of locality and realism put a limit on the strength of the
correlations such that $S_{C}\leq 2.$ \ However, since the four terms in the
Bell inequality are stabilizers of the cluster state, with the last term
having opposite sign, the cluster state can violate this bound up to the
algebraic limit of this expression, i.e. $S_{C}=4$. \ It is a curious fact
that the GHZ state, which is often said to be a maximally-entangled
multi-particle state, cannot violate this inequality. \ Notice that the four
properties for the cluster state include not only four-particle correlations
as in the original GHZ argument, but also three-particle correlations. \
Those terms involving a measurement of $\sigma _{0}$ of photon 1 completely
ignore the state of polarization of that photon. \ Recall that in a GHZ
state this tracing out of one of the qubits leaves the remaining state
completely mixed with only classical correlations between qubits. This is
not the case in the cluster state as its persistency of entanglement allows
for some particles to be ignored before all entanglement is lost. \ Thus
different classes of multiparticle entanglement can exhibit stronger
violations of local realism depending on the nature of the correlations in
the Bell inequality. \

To create the cluster state, we use a method first demonstrated in reference %
\cite{clusterexp}. \ For the experiment, we generate polarization-entangled
photon pairs using type-II parametric down-conversion \cite{Kwiat}. A
UV-laser pulse with a central wavelength of 395nm and a pulse duration of
200 fs makes two passes through a $\beta $-barium borate (BBO) crystal which
emits entangled photons into the forward pair of modes \emph{a }\&\emph{\ b}
and into the backward pair of modes \emph{c }\&\emph{\ d} (Figure 1).
Transversal and longitudinal walk-off effects are erased by compensating
crystals, which exist of a half wave plate (HWP) implementing a 90$%
{{}^\circ}%
$ rotation and an additional BBO crystal. These compensators are placed in
each of the four modes. Final HWPs, one in mode \emph{a} and another in mode
\emph{c}, and the tilt of the compensation crystal allows the generation of
any of the four Bell states. The forward pair of modes \emph{a }\&\emph{\ b}
are coherently superimposed with the backward pair of modes \emph{c }\&\emph{%
\ d} at the two polarizing beamsplitters (PBS) by adjusting the position of
the delay mirror for the UV-pump. The preparation of the cluster state
relies on all of the lowest-order processes which result in the simultaneous
emission of four photons.

Recall that the PBS is a device which transmits horizontally polarized light
and reflects vertically polarized light. \ If two photons enter a PBS from
opposite input ports, they will only emerge separately if their
polarizations are the same in the H/V basis. \ If two-photons enter a PBS
from the same input port, they only emerge separately if they are oppositely
polarized in the H/V basis. \ In the present case, the source was aligned to
produce the Bell state $\left| \phi ^{-}\right\rangle $ into modes \emph{a }%
\&\emph{\ b} and $\left| \phi ^{+}\right\rangle $ into modes \emph{c }\&%
\emph{\ d}. \ If one pair of photons is emitted into modes \emph{a }\&\emph{%
\ b} and another into \emph{c\ }\&\emph{\ d}, then, after the two PBSs, the
four-photon state $\left| HHHH\right\rangle _{1234}-\left| VVVV\right\rangle
_{1234}$ is left, provided the photons emerge into four different output
modes. Emission of two pairs of photons in a single direction occurs with
approximately equal probability, and contributes two more terms to the final
state\ $-\left| HHVV\right\rangle _{1234}$ coming from the first pass and\ $%
\left| VVHH\right\rangle _{1234}$ from the second. \ Provided that all of
these processes are indistinguishable and their relative phases are fixed,
the final state is a coherent superposition of all four terms. \ The
requisite $\pi $-phase shift on the $-\left| HHVV\right\rangle _{1234}$ term
to $+\left| HHVV\right\rangle _{1234}$ was implemented using the HWP in mode
\emph{a}. \ A HWP rotation by an angle, $\theta $, modifies the amplitude of
this term according to the relation $-\cos 2\theta \left| HHVV\right\rangle
_{1234},$ thus a rotation of larger than 45$^{\circ }$ adds the required
phase shift. \ Note that this rotation also changes the amplitudes of the $%
\left| HHHH\right\rangle _{1234}\ $\ and $\left| VVVV\right\rangle _{1234}$
terms by a factor of $\cos \theta $. \ Single-mode fibre-coupled photon
counters were used in modes 1-4 to detect the photons. \ Controlling the
coincidence counting rates from the forward and backward pairs give the
extra degrees of freedom to balance the four amplitudes in the state.

The required expectation values, comprising products of Pauli operators,
were reconstructed from sets of multi-particle polarization correlation
measurements. \ Each of the 48 measurements was performed\ for 600$s$ using
combinations of QWPs and linear polarizers in each of the 4 output modes
(1-4). The Pauli operators $\sigma _{X,Y,Z}$ were measured by projective
polarization measurements, $\left| H/V\right\rangle $ for the $\sigma _{Z}$
operator, $\left| +/-\right\rangle =\frac{1}{\sqrt{2}}\left( \left|
H\right\rangle \pm \left| V\right\rangle \right) $ for the $\sigma _{X}$
operator, and $\left| R/L\right\rangle =\frac{1}{\sqrt{2}}\left( \left|
H\right\rangle \pm i\left| V\right\rangle \right) $ for the $\sigma _{Y}$
operator. For the linear polarization measurements the QWP was set parallel
to the orientation of the polarizer, whereas for the circular polarization
measurements, the QWP was fixed to $+45%
{{}^\circ}%
$ while the polarizer was horizontally- or vertically- oriented. In order to
extract the expectation value, 16 (four-particle correlations) or 8
(three-particle correlations) measurements are required.

Experimental imperfections, including partial distinguishability in the four
relevant four-photon emission processes and phase instabilities, lead to
imperfect correlations which give some coincidence counts even when theory
predicts none. \ For the three-particle correlations, we removed the
polarizer from mode 1. \ However, since the state preparation method was
reliant upon post-selection, four-fold coincidences were still collected. \
Those measurements made without the polarizer show an increase in the
coincidence rate as well as an imbalance most likely due to changes in the
sensitive single-mode spatial filtering.

The four extracted correlations are shown in Figure 2. \ We obtained
positive expectation values of $0.61\pm 0.05$, $0.59\pm 0.04$ and $0.71\pm
0.04$ for the measurements $\sigma _{X}\sigma _{Y}\sigma _{X}\sigma _{Y},$ $%
\sigma _{X}\sigma _{Y}\sigma _{Y}\sigma _{X}$ and $\sigma _{0}\sigma
_{Z}\sigma _{X}\sigma _{X}$, respectively, and the negative value $-0.69\pm
0.04.$ for $\sigma _{0}\sigma _{Z}\sigma _{Y}\sigma _{Y}$. \ Adding these
four correlations together according to the Bell inequality from Eq. 3
results in $S_{C}=2.59\pm 0.08$, where the uncertainty is due to Poisson
counting statistics. \ The threshold for a local realistic modeling of these
correlations is $S_{C}\leq 2$, which our experiment violates by 7$\sigma $.

The remarkable entanglement properties of cluster states can be readily used
for the alternative ``one-way'' model of quantum computing~\cite{briegel},
as was recently demonstrated experimentally~\cite{clusterexp}. \ Different
from the two well-known classes of multiparticle entanglement, GHZ and W
type, the properties of cluster states, such as their robustness against
decoherence and their persistency of entanglement make them practical for
experimental study and interesting for quantum foundations. \ In this
experiment, we have addressed a question of more fundamental rather than
practical interest, namely how the novel family of cluster states can be
used to demonstrate the non-local facets of quantum physics. We investigated
a new kind of Bell inequality based on a GHZ argument for cluster states.
The inequality detects cluster state entanglement optimally, while GHZ
states would not violate the inequality. Our experimentally-produced cluster
violates the inequality by more than 7$\sigma $. Our result demonstrates how
specifically tailored Bell inequalities (e.g. by using specific correlations
of the state) can become a useful tool to tackle the interesting questions
between multiparticle entanglement and quantum nonlocality.

The authors thank V. Scarani and E. Schenck for helpful discussions. \ This
work was supported by ARC Seibersdorf Research GmbH, the Austrian Science
Foundation (FWF), project number SFB 015 P06, NSERC, the European
Commission, contract number IST-2001-38864 (RAMBOQ), and by the Alexander
von Humboldt-Foundation.

\textbf{Figure 1. Experimental setup for the generation of four-photon
cluster states. An ultraviolet laser pulse passes twice through a nonlinear
crystal which is aligned to produce polarization entangled photon pairs on
both the first and second pass. Compensators (Comp) are placed in the modes
\emph{a}, \emph{b}, \emph{c}, and \emph{d} to compensate birefringent
effects and half waveplates (HWP) in mode \emph{a}, \emph{c} to manipulate
the emitted entangled pairs. In each output mode quarter waveplates (QWP)
and polarizers (Pol) are placed to project onto any desired state. \
Including the possibility of double-pair emission and the action of the
polarizing beam-splitters (PBS), the four components of the cluster state
are prepared. The incorrect phase on the HHVV amplitude can easily be
changed by using the HWP in mode \emph{a}. Using these processes and
multiphoton coincidence post-selection, the four-photon cluster state }$%
\left| \phi _{4}\right\rangle $ \textbf{is generated in modes 1-4.}

\textbf{Figure 2 Experimentally extracted polarization correlations. The
cluster-state Bell inequality requires four different polarization
correlations. \ These are extracted from a complete set of 48 four-fold
coincidence measurements. The four-photon correlations, }$\sigma _{X}\sigma
_{Y}\sigma _{X}\sigma _{Y}$ \textbf{and} $\sigma _{X}\sigma _{Y}\sigma
_{Y}\sigma _{X\text{,}}$\textbf{\ are combinations of 16 coincidence rates,
whereas the three-photon correlations, }$\sigma _{0}\sigma _{Z}\sigma
_{X}\sigma _{X}$ \textbf{and} $\sigma _{0}\sigma _{Z}\sigma _{Y}\sigma _{Y}$%
\textbf{, are combinations of 8 coincidence rates. \ Each measurement run
was recorded for 600s. \ A quarter-wave plate and linear polarizer were used
for each polarization projection. \ The polarizer could be completely
removed for those cases where }$\sigma _{0}$\ \textbf{was measured.} \textbf{%
\ The values for the correlations }$\sigma _{X}\sigma _{Y}\sigma _{X}\sigma
_{Y}$, $\sigma _{X}\sigma _{Y}\sigma _{Y}\sigma _{X}$, \textbf{\ }$\sigma
_{0}\sigma _{Z}\sigma _{X}\sigma _{X}$ \textbf{and} $\sigma _{0}\sigma
_{Z}\sigma _{Y}\sigma _{Y}$ \textbf{are }$\left( +0.61\pm 0.05\right) $%
\textbf{,} $\left( +0.59\pm 0.04\right) $\textbf{, }$\left( +0.71\pm
0.04\right) $\textbf{, and }$\left( -0.69\pm 0.04\right) $\textbf{,
respectively. \ Substituting these into the Bell inequality in Eq. 3 yields
a Bell parameter, }$S_{C}\mathbf{=}2.59\pm 0.08$\textbf{, which violates the
local realism threshold by more than 7}$\sigma $\textbf{.}\pagebreak


\begin{thebibliography}{99}
\bibitem{footnote} {One surprising consequence is the existence of a
GHZ-argument for mixed cluster states.}
\end{thebibliography}
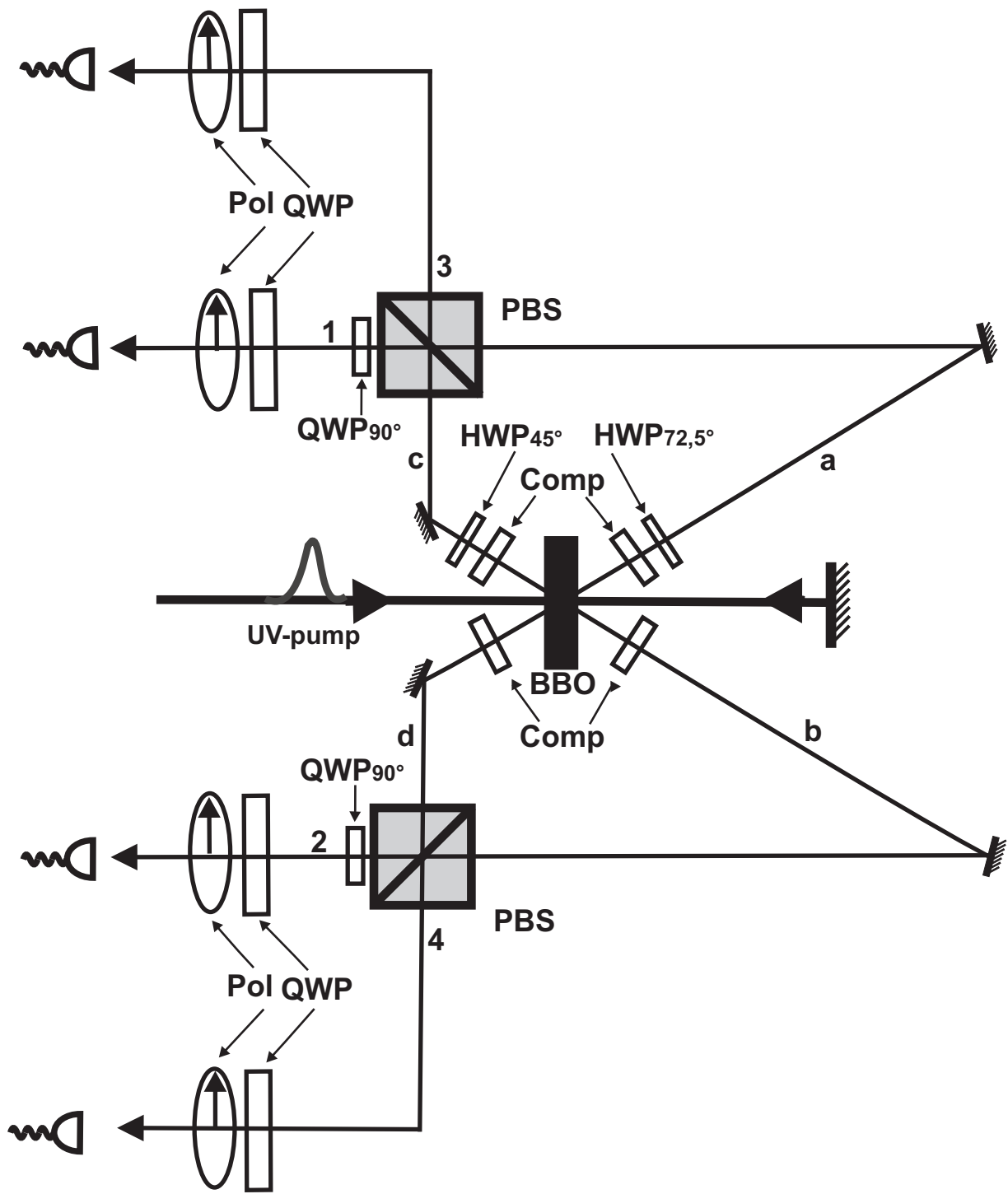
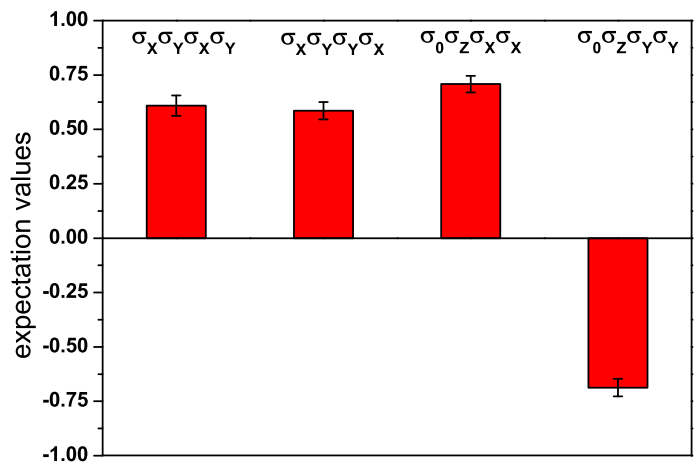
\end{document}